\pdfoutput=1
\documentclass{article}

\usepackage[final]{neurips_2024}

\usepackage[utf8]{inputenc} 
\usepackage[T1]{fontenc}    
\usepackage[hidelinks]{hyperref} 
\usepackage{url}            
\usepackage{booktabs}       
\usepackage{amsfonts}       
\usepackage{nicefrac}       
\usepackage{microtype}      
\usepackage{xcolor}         
\usepackage[hang,flushmargin]{footmisc}
\usepackage{enumitem}
\usepackage{xspace}
\usepackage{arydshln}
\usepackage{multirow}
\usepackage{graphicx}

\usepackage{etoolbox}
\makeatletter
\patchcmd{\@verbatim}
  {\verbatim@font}
  {\verbatim@font\small}
  {}{}
\makeatother

\newcommand{\ignore}[1]{}

\title{Musings About the Future of Search:\\ A Return to the Past?}

\author{Jimmy Lin,$^{1}$ Pankaj Gupta,$^{2}$ Will Horn,$^{2}$  Gilad Mishne$^{2}$ \\[1ex]
$^{1}$ University of Waterloo \qquad $^{2}$ BSL AI
}

\begin{document}
\maketitle

\begin{abstract}
When you have a question, the most effective way to have the question answered is to directly connect with experts on the topic and have a conversation with them.
Prior to the invention of writing, this was the only way.
Although effective, this solution exhibits scalability challenges.
Writing allowed knowledge to be materialized, preserved, and replicated, enabling the development of different technologies over the centuries to connect information seekers with relevant information.
This progression ultimately culminated in the ten-blue-links web search paradigm we're familiar with, just before the recent emergence of generative AI.
However, we often forget that consuming static content is an imperfect solution.
With the advent of large language models, it has become possible to develop a superior experience by allowing users to directly engage with experts.
These interactions can of course satisfy information needs, but expert models can do so much more.
This coming future requires reimagining search.
\end{abstract}

Having questions and seeking answers to them are as old as humanity.
The information science and information retrieval literature broadly refers to this process as information seeking, which is the terminology we adopt here.

The most effective information seeking solution is for the information seeker to directly connect with the best known expert\footnote{{\it The} best expert or a small panel of experts? We're agnostic on this distinction, although for expository purposes writing in plural allows us to avoid the use of gendered pronouns.} on the topic of interest and have a conversation, interacting with the expert until the information need has been satisfied~\citep{Taylor62}.\footnote{In his words:\ ``We assume that man talking with man is the best possible form of communication.''}
In our distant preliterate past, this was the only way.
It is our hypothesis that the future of search is to return to this past, but aided by yet-to-be-developed technologies and millions of large language models (LLMs) that overcome fundamental challenges that have plagued humanity for millennia.

Writing is a relatively late invention in human history, a development that occurred roughly 5000 years ago~\citep{Robinson_1995}.
In preliterate societies, seeking information by finding relevant experts was the only way.
If you had a question, you went to the people in your village who were the most knowledgeable about the topic and simply asked them---or, more accurately, had a conversation with them until your question was answered.

Although effective, this ``ask-the-experts'' approach exhibits two fundamental limitations, both around its scalability:

\begin{enumerate}

\item {\bf The limited bandwidth of experts}, who could practically have only a small number of conversations with information seekers daily.
Quite simply, these experts had other things to do (e.g., hunting, farming, etc.)\ and couldn't spend every waking moment answering questions.
However, as societies developed specialization of roles, these experts evolved to fill the niche of sages or village elders, sanctioned by the community to devote significant time to transmitting knowledge orally.

\item {\bf The problem of finding relevant experts and their willingness to engage}, especially as populations grew larger and became more geographically distributed.
Dunbar's number has been suggested as the cognitive limit of the number of people that an individual can maintain stable social relationships with~\citep{Dunbar92}.
Some limit like this sets the upper bound on the pool of experts one could know, who might also be willing to share their time.
Beyond this limit, some type of referral is likely necessary.

\end{enumerate}

Before the invention of writing, knowledge was locked in the heads of individuals, and ``connecting directly with experts'' was the only way to ``search'', to address information needs.
The invention of writing, however, provided a mechanism for materializing knowledge in forms that became increasingly easier to preserve and replicate.

Worth repeating:\ the invention of writing enabled the convenient materialization of knowledge.

This invention solved the bandwidth problem.
The speed of information dissemination was no longer bounded by the physical limitations associated with verbal transmission by experts.
Instead, they could write down (i.e., materialize) what they knew, whether on clay tablets, tortoise shells, vellum, papyrus, or any other medium---to be consumed by others, across time and space, with a degree of parallelization.\footnote{Also from \cite{Taylor62}: ``Our ideal is to bring two people together through the printed page.''}

The materialization of knowledge solved a number of practical problems, for example:

\begin{itemize}

\item The experts don't have the time or the desire to converse with every information seeker.

\item The experts aren't geographically close to the information seekers.

\item The experts aren't ``temporally'' close to the information seekers.
In fact, the experts may not even be alive anymore!

\end{itemize}

Furthermore, it became possible to replicate materialized knowledge (i.e., make copies), thus leveraging network effects and parallelizing dissemination to a degree that would not have been possible with human experts transmitting knowledge orally.
Centuries ago, replicating materialized knowledge was an expensive proposition, when human scribes had to meticulously make copies by hand.
The invention of the printing press (mad props to Gutenberg) dramatically lowered costs, which steadily declined over time, until it was pushed even lower by the invention of the web (hats off to timbl).
Today, the marginal cost of making another copy of something is near zero.

However, we must recognize that such mechanisms lead to a sub-optimal ``user experience''.
The information seeker cannot engage with a ``dead'' medium (i.e., static content that is unable to support interactions).
Often, the materialized knowledge is not exactly what the information seeker desired, for example, differing in a key assumption or detail.
The information seeker cannot engage in follow-up exchanges to clarify a point, to drill down into detail, etc.
Materialized knowledge captured in a static medium represents at best a proxy.
Nevertheless, this imperfect solution represented a huge advance for civilization.
Writing provided a practical workaround to the first challenge of scalability:\ experts had limited bandwidth, so we materialized their knowledge.

The invention of writing was followed by numerous attempts over the centuries to tackle the second challenge:\ finding the right experts who held the knowledge sought after.
Since humanity now dealt with proxies of knowledge (e.g., scrolls and books), the challenge was reformulated into that of finding the relevant materialized proxies.
Today, we call this the search problem.
Over the centuries, great minds proposed different solutions:
The Great Library of Alexandria had Callimachus' Pinakes, a catalog system that connected scholars with the library's holdings~\citep{El-Abbadi_1990}.
Dewey took a stab at the problem in the 19th century~\citep{dewey} (whose solution remains in prod today).

Fast forward to the web just before the advent of large language models (LLMs):\ search engines can be understood as facilitating a ``connection'' across time and space by surfacing materialized proxies of knowledge (i.e., the ten blue links) that were {\it relevant}, thus potentially satisfying a user's information need (again, hearkening back to~\cite{Taylor62}).
These are webpages, blog posts, videos, papers, etc., the things we ``search for''.
The experience remained sub-optimal because the information seeker was still limited to consuming ``dead'' content.
The first challenge remains unaddressed, as it is still not possible to have conversations with the best experts in the world on a topic of interest.
Thus, consuming materialized proxies of their knowledge (reading the webpages and blog posts they wrote, watching videos they produced, etc.)\ remains the next best option.

But lest we forget that these materialized proxies are at best dancing shadows in Plato's cave.
Can we develop a better solution to the first problem?
For example:

\begin{itemize}

\item Instead of reading Jeff's guide on restaurants in London because you're looking for dining options while on vacation, you just want to talk to Jeff directly and ask him for a suggestion in Southwark.
(But does Jeff have the patience to listen as you describe your idiosyncratic palate and your fondness of high-end sake?)

\item Instead of watching Jane's YouTube tutorial on repairing stucco, you'd like to send her a picture of the damage you're facing and ask her to walk you through repairs with ``turn-by-turn'' instructions calibrated to your skill level.
(Of course, she needs to first listen to you ramble about existing DIY experience and the tools that you have access to.)

\end{itemize}

With LLMs, this becomes possible.
You will be able to converse directly with Jeff or Jane.
Well, not exactly the humans Jeff and Jane, but rather their representative LLMs.
We can call these avatars, agents, digital twins, or any other similar name.
Admittedly, we're still not connecting information seekers to {\it human} experts, but this is probably the best we can do for the foreseeable future.

Economics does not pose a barrier to the massive proliferation of these LLMs.
Once created, the marginal cost of replicating knowledge in this form is primarily dictated by the computation costs of LLM inference, which according to one estimate has been decreasing by 10$\times$ every year~\citep{cost}.
Jevons would predict a commensurate increase in the number of available LLMs and their use~\citep{jevons}.
There is emerging consensus that the future does not lie in ``one model to rule them all''~\citep{one-model}.
Rather, we will witness a proliferation of LLMs---thousands, perhaps even millions.\footnote{These millions of LLMs may very well derive from a relatively small set of model backbones, but adapted (post-training) in myriad ways.}
The continued popularity of open-weight models will ensure this future, barring any unexpected regulatory surprises.

In tomorrow's world populated by millions of LLMs, each with distinct expertise, the first challenge is solved.
Expertise has become abundant, and the bandwidth of experts is only limited by the amount of compute capacity we can muster (a separate challenge).
Interacting with agents, avatars, or whatever we want to call them provides a superior ``user experience'' to consuming static materialized content (books, webpages, papers, blog posts, etc.), which is our only option today.

In fact, user interactions with these LLM experts can go far beyond the answering of questions.
``Find examples of verses in iambic pentameter about love''\ is an information need.
``Compose a poem for my wife about our first date in iambic pentameter''\ seems like much more?
One might argue that underlying every information need is a desire for some sort of action; the literature has given ``the ultimate goal'' of information seeking different names, including what information science researchers have called the ``outcome'' or ``use''~\citep{Wilson99}, and what information retrieval researchers have more recently described as the underlying task~\citep{ShahChirag_etal_CHIIR2023}.
In this case, searching for examples of verses in iambic pentameter fulfills an instructional purpose, to aid in the task of composing poetry.
But here the LLM can directly assist in the task!
Perhaps this is not qualitatively different from inferring the intent to purchase an item (and providing online shopping assistance)\ based on a search query comparing two alternatives.
Nevertheless, technology linking information to action so explicitly has never been available before, and LLMs enable interactions that would be impossible with static content.

However, a world with millions of LLMs completely breaks the working solution to the second challenge.
Search techniques and systems today for connecting users to relevant materialized knowledge do not work for this coming future.
We're no longer searching for documents, webpages, images, and video; it's more than just content created on the fly in a personalized manner by generative AI techniques.
Ultimately, we're looking for potential interactions with relevant experts that can go far beyond simply answering questions.
In fact, the very term ``search'' is perhaps insufficient to describe this future world when LLM experts can ``just do it'', like composing a poem or fixing a bug.
Or perhaps we should reconceive search as not merely seeking information, but finding the right assistant to fulfill my ``wants''.

Either way, search needs to be reimagined in a world populated by millions of expert LLMs (dare we call them AIs?).
Our first step is to start a conversation on what this future might look like.

\bibliography{custom}
\bibliographystyle{abbrvnat}

\end{document}